\documentclass[fleqn,twoside]{article}
\usepackage[headings]{espcrc2}

\readRCS
$Id: espcrc2.tex,v 1.2 2004/02/24 11:22:11 spepping Exp $
\usepackage{graphicx}

\usepackage{epsfig}
\usepackage[figuresright]{rotating}

\newcommand{\AmS}{{\protect\the\textfont2
  A\kern-.1667em\lower.5ex\hbox{M}\kern-.125emS}}

\begin{document}

\begin{titlepage}

\begin{flushright}
PCCF RI 04.11\\
hep-ex/0412046
\end{flushright}

\vspace{1cm}
\begin{center}
\Large\bf Status and expected performance of the LHCb experiment
\end{center}

\vspace{1.0cm}
\begin{center}
{\large Pascal Perret}\\[0.1cm]
{\sl Laboratoire de Physique
        Corpusculaire, Universit\'e Blaise Pascal -- CNRS/IN2P3 \\
        24 avenue des Landais, F 63 177 Aubi\`ere cedex, France }
\end{center}

\vspace{1.2cm}

\begin{center}
{\large {\bf Abstract}}
\end{center}

\vspace{0.3cm}

\begin{quotation}
\noindent
LHCb is a dedicated $b$-physics experiment at the future LHC 
collider. Its construction has started and it will be ready to take 
data from the start of LHC operation, scheduled in 2007, and directly
at its full physics potential. LHCb will benefit from an unprecedented
source of $b$-hadrons, provided by LHC, to improve substantially precision
measurements of CP violation parameters in many different and
complementary channels. 
The detector provides good particle identification, vertexing and has 
an efficient and flexible trigger. Its status and expected performance
are reviewed. 
\end{quotation}

\vspace{1.0cm}

\begin{center} 
{\sl Invited talk at BEACH 2004 \\
6$^{\rm th}$ International Conference on Hyperons, Charm \&
  Beauty Hadrons\\
Chicago, United States, June 27 -- July 3 2004 \\
To appear in the Proceedings}
\end{center}

\vfill
\noindent
PCCF RI 04.11\\
August 2004

\end{titlepage}

\thispagestyle{empty}
\vbox{}
\newpage
 
\setcounter{page}{1}



\title{Status and expected performance of the LHCb experiment}

\author{Pascal Perret \address{Laboratoire de Physique
        Corpusculaire, Universit\'e Blaise Pascal - CNRS/IN2P3, \\
        24 avenue des Landais, F 63 177 Aubi\`ere cedex, France.} 
        on behalf of the LHCb Collaboration 
        }


\runtitle{Status and expected performance of the LHCb experiment}
\runauthor{P. Perret}


\begin{abstract}
LHCb is a dedicated $b$-physics experiment at the future LHC 
collider. Its construction has started and it will be ready to take 
data from the start of LHC operation, scheduled in 2007, and directly
at its full physics potential. LHCb will benefit from an unprecedented
source of $b$-hadrons, provided by LHC, to improve substantially precision
measurements of CP violation parameters in many different and
complementary channels. 
The detector provides good particle identification, vertexing and has 
an efficient and flexible trigger. Its status and expected performance
are reviewed. 

\vspace{1pc}
\end{abstract}

\maketitle

\section{INTRODUCTION}
On one hand the Standard Model (SM) predicts large CP violating
asymmetries for $B$ mesons, in many but often rare decays. On another
hand the LHC collider will provide an unprecedented source of
$b$-hadrons. Benefit will be taken from that by constructing the LHCb
experiment~\cite{proposal}.  
It has been conceived to study CP violation and other
rare phenomena in $b$-hadron decays with very high precision. 
This
should provide a profound understanding of quark flavour physics in the
framework of the Standard Model, and may reveal a sign of physics
beyond.

\begin{figure}[h]
\begin{center}
\epsfig{file=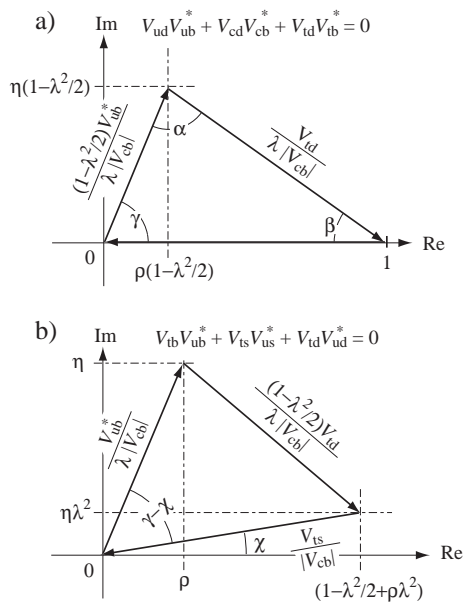, height=8.cm}
\vspace{ -0.8 cm}
\caption{ The two non-squashed unitarity triangles in the Wolfenstein's
  parametrization~\cite{wolf}.} 
\label{fig-triang}
\end{center}
\end{figure}

At LHC energies not only are huge numbers of $B$ mesons produced, 
but there are also produced a large variety of final states involving a
$b$-hadron ($B_{d}$, $B_{u}$, $B_{s}$, $B_{c}$, 
$\Lambda_{b}$, ...). In particular, LHCb will be also capable of
measuring CP violation effects for the first time in  decay modes
involving  $B_{s}$ mesons. The LHCb experiment, thanks to robust
and efficient triggering and particle identification systems, will be
able to exploit this large production to detect and measure pure hadronic
and multi-body final states, and also to constrain the
unitarity triangles of the CKM~\cite{CKM} matrix (Figure~\ref{fig-triang}).

Thanks to the large $b \bar b$ production yield, LHCb will also
have the opportunity to investigate very rare decays of
$b$-mesons~\cite{proposal,reop_tdr,viret}, 
for example, $B^0 \to K^{*0} \gamma$, $B^0 \to K^{*0} \mu\mu$,
$B^0_s \to \mu\mu$, etc.

Complementary measurements will be made of the parameters in different
channels, giving additional cross checks. If new physics is present,
over-constraining the CKM unitarity triangle will be very important in
order to disentangle the SM components from the New Physics.

\section{LHC}

The Large Hadron Collider (LHC) at CERN, Geneva, will collide protons
at a centre-of-mass energy of 14 TeV with a frequency of 40 MHz. The
accelerator will be housed in the 27 km tunnel that has been built for
the LEP collider. LHC is scheduled to start data taking in 2007. The
clear objective is to get the luminosity to $10^{33}$
cm$^{-2}$s$^{-1}$ during 2007 operation. Then the luminosity will be
increased to $10^{34}$ cm$^{-2}$s$^{-1}$, its nominal one, in the next
few years. The construction of the accelerator is 
progressing well. For instance, already more than 300 dipoles are delivered
of a total of 1200. The transfer line is installed, which represents
2.6 km, and first beam is expected in October 2004.

The total cross section at LHC energy is conservatively  assumed to be
$\sim$ 100 mb and the $b \bar b$ total cross section to be $\sim$
500 $\mu$b. Furthermore $b$ and $\bar b$ pairs production are
correlated. They are predominantly produced in the same forward or
backward cone, so that a forward spectrometer like LHCb will have
an acceptance similar to the one of a central detector like ATLAS or
CMS to capture both produced $b$-hadrons.

\section{Satus of LHCb}

LHCb will run at a reduced luminosity of 2 $\times$ $10^{32}$
cm$^{-2}$s$^{-1}$, obtained by controlling the focus of
the beam at the LHCb interaction point. This has been chosen 
to optimise the number of single interactions per
crossing to produce clean events and also to facilitate the triggering
and reconstruction. This permits also to make radiation damage in the forward region
more manageable.   

With this luminosity, LHCb expects about $10^{12}$ $b \bar b$
events per year ($10^7$s). This luminosity is quite moderate compare to the
LHC design one, so it is expected to obtain it from the LHC start-up.

\begin{figure}[bt]
\begin{center}
\epsfig{file=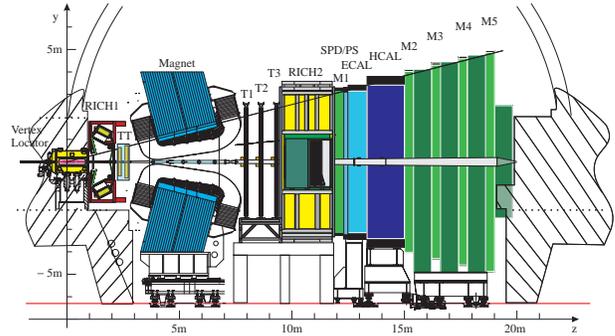,width=80mm}
\vspace{ -0.8 cm}
\caption{ Layout of LHCb showing the
VELO, the two RICH detectors, the tracking stations
TT and T1--T3, the 4 Tm dipole magnet, 
the Scintillating Pad detector (SPD), Preshower (PS), 
Electromagnetic (ECAL) and Hadronic (HCAL) Calorimeters, 
and the muon stations M1--M5. }
\label{fig:lhcb}
\end{center}
\end{figure}

LHCb is a single arm spectrometer covering the range 1.9 $< \, \eta \,
<$ 4.9 (Figure~\ref{fig:lhcb}). It consists of a silicon vertex
detector (VELO) which includes a pile-up system surrounding a beam
pipe, a magnet and a tracking system, two RICH counters, a calorimeter
system and a muon detector. All detector subsystems, except for RICH1, 
are split into two halves that can be separated horizontally for
maintenance and access to the beam pipe.

LHCb has recently undergone a major re-optimisation~\cite{reop_tdr},
which led to a substantially reduced material budget and improved
performances. Its construction has started and is progressing well.

Among the most important features of the LHCb detector are:
an efficient trigger for many $b$ decay topologies;
an efficient particle identification;
an excellent proper time resolution;
a good mass resolution.

 
\subsection{The vertex locator}

The vertex locator (VELO) is installed inside the vacuum tank at the
interaction region~\cite{velo_tdr}. It is made of series of 21
detector stations 
placed along the beam line covering a distance of about 1 m. It
represents 0.23 m$^2$ of silicon, read-out in 170k channels. Each
station consists of two pairs of half-circular Si microstrip
detectors; one with sensors with strips at constant radius ($r$-sensors)
and one with sensors with quasi radial strips ($\phi$-sensors).
The sensors are made from 220 $\mu$m thin silicon. The strip pitch and
length varies as a function of the radial position of the strips
between 37 $\mu$m and 102 $\mu$m. The sensitive area starts at only 8
mm from the beam line. To protect the detectors during LHC beam
injection, they can be retracted to a distance of 3 cm from the beam
line and hence the sensors are placed in Roman pots. They are separated
from the LHC vacuum by a 250 $\mu$m thick Al foil. 

In addition to the 21 VELO stations 
there are two $r$-disks
upstream of the interaction point which make up the Pile-Up System.
It is used in the first trigger level for identifying multiple interaction
events and for measuring the luminosity and track multiplicity.

The VELO provides a resolution on the primary vertex of about 8 $\mu$m
in $(x,y)$ and 44 $\mu$m in $z$. The precision on the impact parameter
is of about 40 $\mu$m and on the proper time resolution it is of about 40
ps$^{-1}$ (for $B_s \to D_s^- \pi^+$).

Construction has started. The vacuum vessel support with its moving
mechanism and the stand are already finished.

\subsection{The tracking system}

In addition to the VELO, the LHCb tracking system consists of one
tracking station before the magnet (TT) and three
tracking stations behind the magnet (T1--T3). It is used to measure
angles and momenta and provides a momentum resolution of $\frac{\delta
  p}{p} \simeq 0.37 \%$. For example the mass resolution is about 14
MeV for $B_s \to D_s K$ decays. The track finding efficiency is
$\sim$ 94 \% for tracks with hits in all tracking stations (for $p$ $>$
10 GeV/$c$).

\subsubsection{The magnet}
To achieve a precision on momentum measurements of better than half a
percent for momenta up to 200 GeV/$c$, the LHC dipole provides integrated
field of 4 Tm~\cite{mag_tdr}. The warm magnet has an aluminium
conductor and is centred at $z$ $\sim$ 5m.
Its weight is 1600 tons and its power
consumption is 4.2 MW. Magnetic field has been introduced between the VELO
and the magnet, $i.e.$ in the region of RICH1 and TT, for Level-1
trigger improvement. The magnetic field will be regularly reversed to
reduce experimental systematic errors in CP violation measurements.

The iron yoke plates and Al coils have been delivered and their
assembly is finished in the LHCb pit. The magnet is moving into its
final position in July 2004 and its commissioning is in Autumn
2004. Field map measurements will be performed in 2004--2005.

\subsubsection{The tracking chambers}

Each tracking station consists of 4 layers. The outer layers (1 and 4)
have vertical readout strips ($x$--layers) to 
measure the track coordinates in the bending plane. The inner layers (2
and 3) are rotated by a stereo angle of +5$^\circ$ and -5$^\circ$
respectively. 
The first station is the Trigger Tracker~\cite{reop_tdr}. It consists
of four planes of silicon microstrip detectors with a pitch of 198
$\mu$m. Strips have lengths of up to 33 cm with a thickness of 500
$\mu$m. They amount to a total surface of approximately 8.3 m$^2$ of
silicon and to 180k readout channels. 
The four layers are arranged in two pairs, with a gap of
30 cm between second and third detection layers. 

The three remaining stations are placed behind the magnet with equal
spacing. Each station consists of an Inner Tracker (IT)~\cite{it_tdr}
close to the beam pipe and  an Outer tracker (OT)~\cite{ot_tdr}
surrounding the IT. The IT uses silicon microstrip detectors with a
pitch of 198 $\mu$m. Strips have lengths of up to 11--22 cm with a
thickness of 320--410 $\mu$m depending on the location. They amount to
a total surface of approximately 4.2 m$^2$  of 
silicon and to  130k readout channels.
The OT is made of 5 mm $\times$ 4.7 m straw tubes, with a fast drift
gas, allowing signal collection in less than 50 ns. It consists of
about 50k readout channels. The serial production has started in 4
different sites. 

\subsection{The RICH}

Two Ring Imaging Cherenkov detectors (RICH1\&2) are used for charged
hadron identification~\cite{rich_tdr}. They are made with three
radiators: Aerogel and C$_4$F$_{10}$ for RICH1, sited upstream of the
magnet, which has an angular coverage 25--300 mrad and CF$_4$ for RICH2
which has an angular coverage 15-120 mrad. They provide greater than 3
$\sigma$ $\pi$/$K$ separation over the momentum interval 3 $<$ $p$ $<$ 80
GeV/$c$. 

The photodetectors used are 1024-pixel LHC-speed Hybrid Photodiodes
(HPD) developed by LHCb and currently under fabrication. The
construction of RICH2 is advancing. The space frame, the
exit/entrance windows, the mirror support are ready.

\subsection{The calorimeter system}

The calorimeter system consists of 4 subdetectors~\cite{cal_tdr}. A
Scintillating Pad Detector (SPD) to distinguish charged particles from
photons is followed by a 15 mm lead wall (2.5 X$_0$) and a Preshower
(PS) made with the same technology. The detector elements are 15 mm
thick scintillator pads. A groove in the scintillator holds the
helicoidal WLS fiber which collects the scintillation light. The light
from both WLS fiber ends is sent by long clear fibers to 64-anode
photomultipliers tubes  that are located above or below the
detector. Just after is the Electromagnetic Calorimeter (ECAL) based
on the Pb/scintillator Shashlik type (25 X$_0$, 1.1 $\lambda_{\rm
  I}$), followed by the Hadronic Calorimeter (HCAL) based on the
iron/scintillator tile type (5.6 $\lambda_{\rm I}$). SPD, PS and ECAL
are made of 5962 channels while HCAL has 1468 channels.

They provide electron, photon and hadron (including $\pi^0$)
identification and play a central role in the Level-0 trigger. Their
readout is performed every 25 ns. The efficiency to identify an
electron is expected to be $\varepsilon$($e \to e$) = 95\% for a
misidentification rate $\varepsilon$($\pi \to e$) = 0.7\%.
In a testbeam, an energy resolution of 
$\frac{\sigma E}{E} = \frac{8.3 \%}{\sqrt{E}} \oplus 1.5 \%$ for the ECAL and
$\frac{\sigma E}{E} = \frac{75 \%}{\sqrt{E}} \oplus 15 \%$ for the HCAL has been measured. 

The construction is progressing well. 30\% of the PS/SPD modules is
completed and the assembly of the supermodules starts in September. 
All the ECAL modules are ready and the installation in the pit will
start in November 2004, while 70\% of the HCAL modules is also
completed and the installation will start by end of 2004.

\subsection{The muon system}

The muon system consists of 5 stations of 1380 Multi Wire Proportional
Chambers~\cite{MuonTDR}. M1 in front of the calorimeter system and
M2--M5 behind it, interleaved with iron shielding plates. M1 is
used mainly to improve the momentum resolution in the first trigger
level. It consists of two layers of MWPC, while the other four stations
are made from four layers. This represents a total surface of
approximately 435 m$^2$, readout in 26k channels and a hadron absorber
thickness of 20 $\lambda_{\rm I}$. 

The system provides efficient muon identification, typically 94\%
for a pion misidentification rate below 1\%. Highest $p_{\rm T}$ muons
are selected and used for Level-0 trigger.
The production has started in various sites, and the muon filter
chariot (2000 tons Fe) is being delivered.

\subsection{The trigger}

The LHCb trigger is dedicated to select $b$ decays of interest. It is
subdivided in three trigger levels, called Level-0, Level-1 and
HLT~\cite{trig_tdr}.  
The aim of Level-0 is to reduce the LHC beam crossing rate of 40 MHz
which contains about 10 MHz of crossings with visible $pp$ interactions
at the LHCb luminosity, to a rate of about 1 MHz at which in principle
all sub-systems could be used for deriving a trigger decision. It
preselects the ``highest $p_{\rm T}$'' ($>$1--2 GeV/$c$) muon tracks and
highest $E_{\rm T}$ calorimeter clusters ($e$, $\gamma$, $\pi^0$ and
hadrons) which information is collected by the Level-0 Decision Unit
(L0DU) to select events. With simple arithmetic, L0DU combines all
signatures into one decision per crossing. Events can be rejected
based on global event variables such as the total transverse energy
deposited in HCAL, the charged track multiplicities, or the number of
interactions as reconstructed by the Pile-Up system. To be fast it is
implemented on hardware.

The Level-1 algorithm will be implemented on a commodity processor
farm, which is shared between Level-1, HLT and offline reconstruction
algorithms. The Level-1 algorithm uses the information from Level-0,
VELO and TT and its implementation is easily scalable to include
other parts of the detector. Events are selected based on tracks with a large
$p_{\rm T}$ and a significant impact parameter to the primary
vertex. The maximum Level-1 output rate has been fixed to 40 kHz, at
which rate full event building is performed. 

The HLT algorithm starts with a pre-trigger which aims at confirming
the Level-1 decision with better resolution, followed by selection
algorithms dedicated to either select specific final states, or
generic cuts to enrich the $b$-content of the events written to storage
at a target rate of about 200 Hz.

The efficiency achieved by the Level-0 and Level-1 triggers varies
between 30\% and 60\% depending on the final states, while the HLT
efficiency will reach nearly 100\%.

\section{EXPECTED PERFORMANCE}

\subsection{Simulation}

The LHCb collaboration uses the PYTHIA 6.2 physics generator in
conjunction with QQ and GEANT3. The track multiplicity has been tuned
to reproduced CDF and UA5 data and the model includes the description
of multiple parton-parton interaction with varying impact parameter.
Multiple $pp$ interactions in the same bunch crossing are also
included. The response of the detector is simulated in a realistic way
including noise and spill-over effects. A complete description of
material from TDRs is used and individual detector responses are tuned
on test beam results. A complete pattern reconstruction in
reconstruction is performed. Reconstruction and selection algorithms
do not make use of the true Monte Carlo information at any stage.

In 2004, EvtGen and GEANT4 are used to generate events.
About 180 million events are being simulated and analysed in a
distributed way (GRID). 

\subsection{Physics}

About 67 million events were produced in 2003 with about 10 million 
$b \bar b$ events, and they are used to evaluate the expected physics
performance quoted here~\cite{reop_tdr}. CKM angles are defined on
Figure~\ref{fig-triang}. Results are given for one year of LHCb data
taking $i.e.$ 2 fb$^{-1}$.

\subsubsection{Flavour tagging}

The identification of the initial flavour of reconstructed $B^0_d$
and $B^0_s$ mesons is often necessary. Different algorithms have
been developed using same-side kaon tag, opposite-side kaon tag,
lepton tag or vertex charge. Combining them, they typically provide an
effective tagging efficiency of around 4\% for  $B^0_d$ and 6\%
for  $B^0_s$ channels. 

\subsubsection{\boldmath $\beta$ from $B^0_d \to J/\psi K^0_S$}

The $B^0_d \to J/\psi K^0_S$ mode is the ``gold-plated'' channel
explored at $B$-factories. Precision measurement of this parameter is
very important. The CKM angle $\beta$ is extracted from a fit to the
time dependent asymmetry:

${\cal A^{\rm CP}} = {\cal A^{\rm dir}}
\cos(\Delta m_d t) + {\cal A^{\rm mix}} \sin(\Delta m_d t)$ 

\noindent 
where ${\cal A^{\rm dir}}$ is expected to be equal to 0 and ${\cal
  A^{\rm mix}}$ to sin2$\beta$ in the SM. LHCb data will bring a lot
  of statistics to this channel, which can be used to look into higher
  order effects and also to fit ${\cal A^{\rm dir}}$. LHCb will collect 
240k events in one year, which will obtain a statistical
  precision on sin2$\beta$ of around $\pm$ 0.02.

\subsubsection{\boldmath $ \Delta m_s$ from $B^0_s \to
  D_s^-(KK\pi)\pi^+$}

LHCb will fully reconstruct this decay to measure the $B_s^0$
oscillation frequency. It will benefit from excellent momentum
resolution (a mass resolution $<$ 14 MeV/$c^2$), decay length resolution
($\sim$ 200 $\mu$m),  proper time resolution ($\sim$ 40 fs) and an
effective tagging efficiency close to 9\%. In one year, LHCb will
collect 80k events and will be able to observe greater than 5$\sigma$
oscillation signal if $ \Delta m_s <$ 68 ps$^{-1}$, which is well
beyond the SM expectation (14.8-26 ps$^{-1}$).
Once a $ B_s \overline{B_s}$ oscillation signal is seen, its
frequency is precisely determined, about $\pm$0.01 ps$^{-1}$.

\subsubsection{\boldmath $\chi$ from $B_s \to J/\psi \phi$}

The decay $B_s \to J/\psi \phi$ is the $B_s$ counterpart of
the ``golden'' mode  $B_d \to J/\psi K_S$ and is particularly
interesting as it gives access to the weak mixing phase of the  
$B_s$. This analysis is complicated by the fact that the final state is
made by two vector mesons and so two orbital momentum states can
occur. Therefore an angular analysis of the decay final states is
needed to separate CP-even and CP-odd contributions. Experimentally,
good proper-time resolution is essential. This is the case in LHCb
($<$38 fs) and feasible thanks to the large sample collected
by reconstructing $\phi$ in $K^+K^-$ and $J/\psi$ in $e^+e^-$ or
in $\mu^+ \mu^-$,
120k events in one year. The attainable precision for sin2$\chi$ is of
the order of $\pm$0.06 and for $\Delta \Gamma_{\rm s}/\Gamma_{\rm s}$ is
$\pm$0.02 (for $\rm \Delta m_s$=20 ps$^{-1}$). 
 
In the SM, the expected asymmetry is very small (sin2$\chi$ $\sim$
0.04), so it offers a sensitive probe for CP violating contributions
beyond the SM.

\subsubsection{\boldmath $\gamma$ from $B^0_d \to \pi^+\pi^-$ and
  $B^0_s \to K^+K^-$} 

In both decays, large $b \to d (s)$ penguin contributions are
present in addition to the  $ b \to u$ tree transition. The
time-dependent decay asymmetries 
${\cal A^{\rm CP}}= {\cal A^{\rm dir}}
\cos(\Delta m_d t) + {\cal A^{\rm mix}} \sin(\Delta m_d t)$ 
can be measured in both channels; this provides four observables which
can be parametrized with seven parameters. By exploiting the U-spin
symmetry of the strong interaction to relate observables in both
decays~\cite{Fleischer} and measurements of $\chi$ from $B_s \to
J/\psi \phi$ and of $\beta$ from $B_d \to J/\psi K^0_S$, these can
be reduced to three parameters and $\gamma$ can easily be
extracted. Particle identification and good mass resolution are vital
for this measurement. In one year, LHCb will collect around 26k of
$B^0_d \to \pi^+\pi^-$ and 37k of $B^0_s \to K^+K^-$ and will have
a sensitivity to the $\gamma$ angle of 4--6 degres.

\subsubsection{\boldmath $\gamma$ from $B^0_d \to
  \overline{D^0}K^{*0}$, $D^0K^{*0}$} 

The simultaneous measurement of the rates for the decays
$B^0_d \to \overline{D^0}(K^+ \pi^-)K^{*0}$, $B^0_d \to
D^0_{CP}(K^+K^-)K^{*0}$, $B^0_d \to D^0(\pi^+ K^-)K^{*0}$
and their CP conjugates, where $K^{*0} \to K^+ \pi^-$, allows the CKM
angle $\gamma$ to be extracted without the need of flavour tagging or
proper-time determination~\cite{gwd}. Since branching ratios are very
small, high statistics and good particle identification are necessary
for this measurement. The LHCb sensitivity to the $\gamma$ angle is of
7--8 degres in one year of data taking.  

\subsubsection{\boldmath $\gamma$ from $B^0_s \to D^\pm_s K^\mp$}

The decay $B^0_s \to D^\pm_s K^\mp$ and its charge conjugate can
proceed through two tree  decay diagrams, the interference of which
gives access to the phase $\gamma - 2 \chi$, and hence to the CKM
angle $\gamma$ if $2 \chi$ is determined otherwise ($e.g.$ with 
$B_s \to J/\psi \phi$). Four time-dependent decay rates are
measured. Particle identification and good mass resolution are very
important; the signal is dominated by  background from the $B^0_s
\to D^-_s \pi^+$ decays, which has a branching fraction 12 times
higher than the signal channel.
LHCb will record about 5.4k events in one year, giving a sensitivity
of 14--15 degres in $\gamma$.

\section{CONCLUSION}

LHCb will perform a study of CP violation with unprecedented precision
in many different and complementary channels, having access to all
$b$-hadron species. Some of the measurements
will be more and some less sensitive to new physics contributions
allowing us to disentangle the effects and to have a comprehensive
understanding. 

The detector provides excellent particle identification, vertex,
momentum and proper decay time resolution and has
an efficient, flexible and robust trigger. Construction has started and is
progressing well. LHCb will be ready for data-taking at
the LHC start-up in 2007 with its nominal luminosity.
It will provide a sensitive test of the Standard
Model and physics beyond.

\end{document}